\documentclass[%
reprint,
superscriptaddress,
 amsmath,
 amssymb,
 aps,
]{revtex4-2}

\usepackage{graphicx}
\graphicspath{{Final_plots/Fig1/}{Final_plots/Fig2/}{Final_plots/Fig3/}{Final_plots/Fig4/}{Final_plots/Appendix}}
\usepackage{comment}
\usepackage{dcolumn}
\usepackage{bm}
\usepackage{hyperref}
\newcommand{\ket}[1]{\left|{#1}\right\rangle}

\usepackage{xfrac}
\usepackage{siunitx}
\usepackage[T1]{fontenc} 
\usepackage[dvipsnames]{xcolor}
\colorlet{ForestGreen}{Black}
\begin{document}

\preprint{APS/123-QED}

\title{Efficient Detection of Statistical RF Fields with a Quantum Sensor}

\author{Rouven Maier}
 \thanks{These authors contributed equally to this work.}
 \affiliation{3rd Institute of Physics, University of Stuttgart, 70569 Stuttgart, Germany.}  
 \affiliation{Max Planck Institute for Solid State Research, 70569 Stuttgart, Germany.}

\author{Cheng-I Ho}
 \thanks{These authors contributed equally to this work.}
 \affiliation{3rd Institute of Physics, University of Stuttgart, 70569 Stuttgart, Germany.}  
 \affiliation{Center for Integrated Quantum Science and Technology, 70569 Stuttgart, Germany.}

\author{Hitoshi Sumiya}
 \affiliation{Advanced Materials Laboratory, Sumitomo Electric Industries Ltd., Itami, Hyougo 664-0016, Japan.}

\author{Shinobu Onoda}
 \affiliation{Takasaki Institute for Advance Quantum Science, 1233 Watanuki, Takasaki, Gunma 370-1292, Japan.}

\author{Junichi Isoya}
 \affiliation{Faculty of Pure and Applied Sciences, University of Tsukuba, Tsukuba, Ibaraki 305-8550, Japan.}  

\author{Vadim Vorobyov}
\email[]{v.vorobyov@pi3.uni-stuttgart.de}
 \affiliation{3rd Institute of Physics, University of Stuttgart, 70569 Stuttgart, Germany.}  
 \affiliation{Center for Integrated Quantum Science and Technology, 70569 Stuttgart, Germany.}

\author{J\"org Wrachtrup}
 \affiliation{3rd Institute of Physics, University of Stuttgart, 70569 Stuttgart, Germany.}
 \affiliation{Max Planck Institute for Solid State Research, 70569 Stuttgart, Germany.}
 \affiliation{Center for Integrated Quantum Science and Technology, 70569 Stuttgart, Germany.}

\begin{abstract}
Nuclear magnetic resonance (NMR) spectroscopy is widely used in fields ranging from chemistry, materials science to neuroscience.
Nanoscale NMR spectroscopy using Nitrogen-vacancy (NV) centers in diamond has emerged as a promising platform due to an unprecedented sensitivity down to the single spin level.
At the nanoscale, high nuclear spin polarization through spin fluctuations (statistical polarization) far outweighs thermal polarization.
However, until now efficient NMR detection using coherent averaging techniques could not be applied to the detection of statistical polarization, leading to long measurement times.
Here we present two protocols to enable coherent averaging of statistically oscillating signals through rectification.
We demonstrate these protocols on an artificial radiofrequency signal detected with a single NV center at \SI{2.7}{T}.
The signal-to-noise scaling with number of measurements $N$ increases from $N^{0.5}$ to $N^1$, improving the measurement time significantly.
The relevance of rectification for the detection of statistically polarized nuclear spins using ensembles of NV centers is outlined, paving the way for efficient nanoscale NMR spectroscopy.

\end{abstract}

\maketitle

\begin{figure}
    \includegraphics{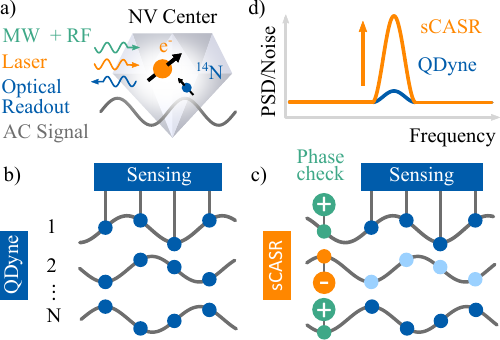}
    \caption{\label{fig:schematic} \textbf{Rectification of Statistical RF Signals.} 
    \textbf{a)} A single Nitrogen-vacancy (NV) center in diamond as quantum sensor to detect an external oscillating (AC) signal.
    \textbf{b)} Classical quantum heterodyne (QDyne) experiment with sequential measurements of the target signal.
    \textbf{c)} Statistical coherently averaged synchronized readout (sCASR) rectification protocol.  
    The initial phase of the target signal is detected through a single detection block.
    The time traces can be coherently averaged based on the detected initial phase.
    \textbf{d)} Normalized power-spectral density (PSD) is drastically improved through sCASR. 
    }
\end{figure}

Quantum sensing of oscillating (AC) signals has gained increasing attention in applied and fundamental science. 
The applications span from radio frequency (RF) \cite{chen2023quantum, Backes_2024} and microwave (MW) detection \cite{joas2017quantum, shao2016wide, meinel2021heterodyne, jing2020atomic, Wang:2022aa} including nuclear magnetic resonance (NMR) \cite{smits2019two,liu2022surface} and electron paramagnetic resonance (EPR) \cite{dwyer2022probing, findler2024detecting, pinto2020readout, kong2020kilohertz}, to high-energy physics such as the search for dark matter \cite{sushkovDark2023}, precision mass measurements of matter and antimatter \cite{ulmerAntiprotontoproton2015}, and tests of Charge, Parity and Time reversal symmetry (CPT) as well as Lorentz invariance violation \cite{brownNewLimitLorentz2010}.
NMR with quantum sensors, in particular, is widely explored, demonstrating an improved detection sensitivity by more than ten orders of magnitude \cite{allertAdvances2022, loretzNanoscale2014, staudacherNMR2013} at room temperature and opening a gateway for magnetic resonance imaging (MRI) of molecules at the nanoscale \cite{rugarProtonNMR2015, lovchinsky2016nuclear}.
In contrast to conventional NMR with high magnetic fields and large sample volumes, where thermal polarization dominates, statistical polarization is the primary source of target spin magnetization in nanoscale NMR. 
It scales as $\mu /\sqrt{n_{sp}}$, where $n_{sp}$ is the number of nuclear spins, and $\mu$ is nuclear magnetic moment \cite{mamin2003detection}. 
Typically, NMR detection with quantum sensors works with statistical polarization and relies on correlation protocols, resembling variations of electron spin echo envelope modulation (ESEEM) \cite{laraouiCorrelation2013,lovchinsky2016} as well as electron nuclear double resonance (ENDOR) type of protocols \cite{maminNMR2013,aslam2017}. 
Here the frequency resolution is limited by the sensor or memory lifetime. 
However, sensitivity is suboptimal due to the time overhead for data sampling. 
Although chemical shift resolution is achievable at the nanoscale, 1D scans require averaging times of days to weeks, making 2D and higher-dimensional NMR studies infeasible \cite{staudacherNMR2013}.
Long measurement times also limit the application of nanoscale NMR in dynamic and in vivo environments as well as measurements under ambient conditions.
Sequential measurement protocols, such as quantum heterodyne (QDyne) \cite{schmittQdyne2017, Boss.2017} and coherently averaged synchronized readout (CASR) \cite{glennCASR2018} enable sensor-unlimited spectral resolution and efficient signal acquisition of radio frequency \cite{Boss.2017,schmittQdyne2017}, microwave \cite{meinelHeterodyne2021, Yin.03122024} and NMR signals \cite{Meinel.2023, Briegel.2025}, with frequency uncertainty scaling as $T^{-3/2}$.  
The methods are characterized by a  significant reduction of the measurement time down to minutes. 
However, their application so far has been limited to the detection of the weaker thermal polarization due to the requirement of phase-coherent target signals \cite{grafensteinCoherent2025}. 
The NMR signal could be substantially increased by using hyperpolarization techniques, at the cost of technical complexity, and limited sample types \cite{Arunkumar.2021,Bucher.2020,Briegel.2025}.
This enabled overcoming the frequency resolution limit caused by spin diffusion, which is encountered in the detection of statistically polarized nanoscale NMR of non-immobilized samples \cite{schwartzBlueprint2019}. 
However, the amount of achieved hyper-polarization is still significantly lower than the available statistical polarization at the nanoscale \cite{grafensteinCoherent2025}.
Till now, the combination of efficient sequential detection through coherent averaging and the detection of statistically polarized target spins at the nanoscale is still missing.
One possible solution is to rectify the signals from statistical polarization to enable coherent signal detection, as has been demonstrated by efficient classical sensors such as magnetic resonance force microscope (MRFM) \cite{peddibhotlaMRFM2013}.

In this work, we establish a protocol for efficient coherent averaging of NMR signals of statistically polarized nuclear spin ensembles through rectification.
To this end, we employ a long-lived ancillary memory qubit adjacent to a quantum sensor, which is capable of storing the phase information of the statistical signal.
Two approaches are shown: a classical feed-forwarding following an efficient projective measurement of the quantum sensor and a coherent feedforward operation (Fig. \ref{fig:schematic}). 
We perform measurements on an artificially created classical RF signal with randomized initial phase and test the applicability of both protocols to potential experimental settings in nanoscale NMR. 
Our results facilitate the detection of AC signals with statistical phases with advanced detection methods such as coherently averaged NMR \cite{glennCASR2018} and electrical current measurements \cite{Garsi.2024}.

\begin{figure}
    \includegraphics{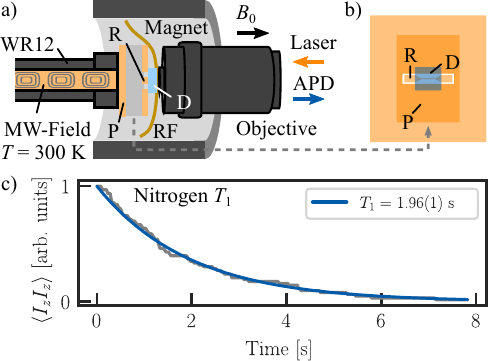}
    \caption{\label{fig:setup} \textbf{Setup and Operation.} 
    \textbf{a)} Experimental setup with home-built confocal microscope at room temperature and the NV diamond located in a superconducting vector magnet at a magnetic field of \SI{2.7}{T}.
    Microwave (MW) control is fed to the sample through a rectangular waveguide (WR 12). 
    \textbf{b)} Spin control of the electron spin in the diamond (D) at \SI{73}{GHz} is achieved through a bow-tie resonator (R), and a patch antenna (P).
    \textbf{c)} Memory lifetime $T_1 = \SI{1.89}{s}$ of the nitrogen spin under constant readout.
    }
\end{figure}

\textit{Results and Discussion.---}To detect the power-spectral-density (PSD) of a signal of interest $s(t)$ (e.g. an oscillating NMR signal) on top of noise $x(t)$, multiple experimental traces $M_i(t) = s_i(t) + x_i(t)$ are typically averaged $N$-times to increase the signal-to-noise ratio (SNR).
In the case of a non-synchronized signal, i.e. when the initial phase of the signal is random for each experimental run $i$, direct averaging of the time trace leads to a vanishing signal, as $\langle s_i \rangle = 0$.
Therefore, the individual power spectra $|\mathcal{F}(M_i)|^2$ are averaged, leading to an improvement of the SNR $\propto \sqrt{N}$.
On the other hand, if the signal is coherent with a predefined phase, direct averaging of the time traces becomes possible, as $\langle s_i \rangle \neq 0$.
Coherent averaging leads to an advantageous SNR scaling $\propto N$ in the PSD (see Supplementary Note 1 for further details).
It is therefore beneficial to convert an incoherent signal into a coherent one, achieved here through \textit{rectification}, to improve the SNR scaling (Fig. \ref{fig:schematic} d).
The rectification relies on two steps.
First, the initial phase $\phi_i$ of the signal in each timetrace is detected to gather information for the rectification step.
Second, the time trace is inverted $M_i \rightarrow - M_i$ if $\phi_i > \pi$, effectively shifting $\phi_i$ by $\pi$.
As a result, all time traces have an effective phase of $\phi_i \in [0, \pi]$ and can be averaged directly.
In our implementation, a long-lived quantum memory is necessary to store the phase information $\phi_i$ and to rectify the time traces.

\begin{figure*}
    \includegraphics{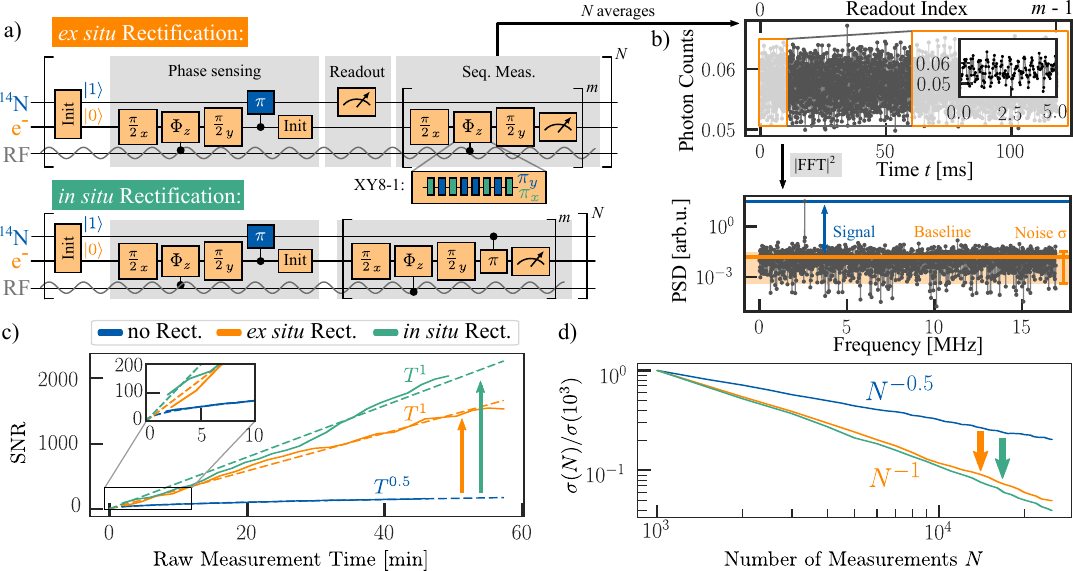}
    \caption{\label{fig:rectification} \textbf{Experimental Detection of Rectified RF Signals.} 
    \textbf{a)} Rectification protocols.
    After initialization, a binary phase information is detected by the electron spin through a XY8-1 block.
    This information is stored on the quantum memory through a controlled $\pi$ rotation of the nitrogen memory spin, before the electron spin is reinitialized.
    In the \textit{ex situ} rectification protocol this information is directly read out via single-shot readout (SSR), before the oscillating time trace is acquired through a sequential sensing and readout block, where $m = 4000$.
    In the \textit{in situ} rectification protocol, a controlled $\pi$ rotation of the electron spin is applied before the readout during the sequential measurement block, inverting the oscillating time trace based on the stored phase information.
    \textbf{b)} Rectified average photon time trace detected by the sequential measurement block ($m$ points).
    The time trace is transformed into the power-spectral-density (PSD) via |FFT|$^2$.
    The signal (blue line) is clearly visible above the noise $\sigma$ (orange errorbar) of the baseline (orange line).
    \textbf{c)} SNR scaling with averaging time $T$ increases from $T^{0.5}$ for non-rectified classical QDyne to $T^1$ for both rectified methods.
    Linear and square root fits are indicated by dashed lines.
    \textbf{d)} Noise scaling with number of measurements $N$ improves from $N^{-0.5}$ in the classical QDyne to $N^{-1}$ in the rectified measurements. For visualization purposes the noise $\sigma_\mathrm{P_{noise}}(N)$ is normalized by $\sigma_\mathrm{P_{noise}}(N=10^3)$.
    }
\end{figure*}

Experimentally, we demonstrate the rectification of a statistical RF signal on a single NV center in diamond; however the protocol is equally applicable to other types of solid state spin sensors possessing a stable ancillary memory qubit.
The adjacent $^{14}$N forms the quantum memory for the rectification protocols.
To increase the longitudinal relaxation time $T_1$ of the memory qubit, we are operating our setup at the maximum available magnetic field of $B_0 = \SI{2.72}{T}$ ($T_1 \propto B_0^2$) \cite{Neumann.2010}.
Under these conditions, control of the sensor spin becomes challenging because of the limited availability of commercial microwave equipment in the corresponding frequency range of \SI{73}{GHz}.
Here, it is achieved by using a bow-tie resonator close to the NV center, allowing robust spin control for sensing protocols.
For a more detailed description of the setup and the resonator design, see Supplementary Note 2 and 3.
At a magnetic field of \SI{2.7}{T}, the lifetime of the quantum memory under perturbations induced by the sensor spin manipulation reaches $T_1 = \SI{1.89}{s}$, allowing Fourier-limited frequency resolutions of $\sim \SI{1}{Hz}$ in the RF detection protocols (Fig. \ref{fig:setup} c).
In the context of NMR spectroscopy, this would be enough to reach a Larmor frequency resolution of \SI{0.01}{ppm} for hydrogen spins at \SI{2.7}{T}, which would be sufficient to resolve chemical shifts ($\sim$ \SI{1}{ppm}) and $J$-couplings ($\sim$ \SI{1}{Hz}) in NMR spectra.

Based on the long-lived quantum memory, we introduce two experimental protocols for the rectification of statistically oscillating signals (Fig. \ref{fig:rectification}a).
In both protocols, the memory spin and the electron spin are initialized, before the random initial phase $\phi_i$ of the target signal is probed by the electron spin via a dynamical decoupling (DD) sequence, acquiring the phase $\theta = \alpha \cos{(\phi_i)}$, where $\alpha$ is the interaction strength between the target signal and the electron spin.
A second MW $\pi$/2 pulse is applied, storing the phase information in the polarization along the $z$ axis of the electron spin.
Finally, this information is transferred onto the memory spin by a controlled $\pi$ rotation of the nitrogen spin.
This results in a binary mapping of the continuous random phase $\phi_i$ of the target signal onto the nitrogen hyperfine levels $\ket{0}$ and $\ket{1}$.
The fidelity $F$ of this mapping is limited by the quantum shot-noise of the state projection into $\ket{0}$ and $\ket{1}$. 
It is optimized in our experiment to reach $F \sim \SI{89}{\percent}$ (Supplementary Note 4).
For the \textit{ex situ} rectification protocol, the memory spin state is directly read out via single-shot readout (SSR), providing the phase information for the rectification in the post-processing.
Next, the target signal is sequentially probed with $m$ DD sensing blocks and read out to extract the oscillating time trace $M_i$ in $m$ points.
In the post-processing, the extracted time traces $M_i$ can be added or subtracted coherently conditioned on the result of the readout of the memory spin, which contains information about the initial phase $\phi_i$.
For the \textit{in situ} rectification protocol, instead of reading out the memory spin directly, a controlled $\pi$ rotation of the electron spin is added before each readout of the sequential measurement to invert the traces based on the phase information previously stored in the memory spin, thus producing coherent time traces \textit{in situ}.
This allows the direct averaging of the oscillating time traces, without additional post-processing for the rectification of the signal.
An example of the averaged time trace is shown in the upper panel of Fig. \ref{fig:rectification}b, where the oscillations of the target signal are clearly visible.
The power spectral density is obtained via the absolute square of the Fourier transform of the time trace (Fig. \ref{fig:rectification}b, lower panel).
The SNR is defined as the difference of the peak-height and the baseline over the root-mean-square of the baseline, as usual in the NMR community.
For this demonstration, an applied RF signal with frequency $f = \SI{166.666}{kHz}$ and an amplitude of $B_\mathrm{signal} = \SI{664(15)}{nT}$ at the position of the NV center is used as target.
The obtained frequency in the PSD is shifted from the frequency of the applied RF due to undersampling, but can be easily transformed back using the sampling frequency.
In all our measurements the spectral resolution is Fourier-limited by the measurement time of the sequential measurement block of $\SI{130}{ms}$.
A comparison of the obtained SNR with and without rectification protocols is shown in Figure \ref{fig:rectification} c for up to $N = 25000$ averages.
Especially for longer measurement times, a clear improvement of the SNR by about one order of magnitude through signal rectification can be seen, demonstrating the advantage of the rectification protocols.
It is worth noting that the measurement times are not significantly increased through the experimental overhead of the rectification protocols.
For a detailed description of the experimental parameters see Supplementary Note 5.
As expected, the SNR-scaling with measurement time improves from $T^{0.5}$ in the case of incoherent averaging without rectification to $T^{1}$ with rectification, leading to a large improvement for longer measurement times. 
The reason for this is the scaling of the noise $\sigma_\mathrm{P, noise}$, which improves from $T^{-0.5}$ to $T^{-1}$ (Fig. \ref{fig:rectification}d), while the total signal remains constant with measurement time.
For a detailed derivation of the SNR scaling under these experimental conditions, see Supplementary Note 6.
The lower slope of the SNR in the \textit{ex situ} rectification compared to the \textit{in situ} rectification are most likely due to pulse imperfections in the experimental realization of the protocol.
Both rectification protocols can be used to rectify statistical AC signals and enable coherent time trace averaging.
The \textit{ex situ} protocol is well suited for experimental conditions, where the lifetime of the memory is limited and the lifetime of the signal is long.
As the information stored on the memory spin is already fully extracted before the oscillating time trace is recorded, erasure of the memory during the sensing block is irrelevant for the rectification process.
Since the direct readout of the memory is a relatively long process (\SI{12}{ms}), this comes at the cost of an increased sequence time, which can cause problems if the lifetime of the target signal is limited, e.g. in diffusion limited NMR sensing.
In contrast, the \textit{in situ} rectification is well-matched for situations, where the lifetime of the memory is long and the lifetime of the signal is limited, due to lower sequence overhead through rectification compared to the \textit{ex situ} rectification.
Also, the absence of an active readout during the rectification process makes this method inherently suitable for the application in ensemble NV experiments.

\begin{figure}
    \includegraphics{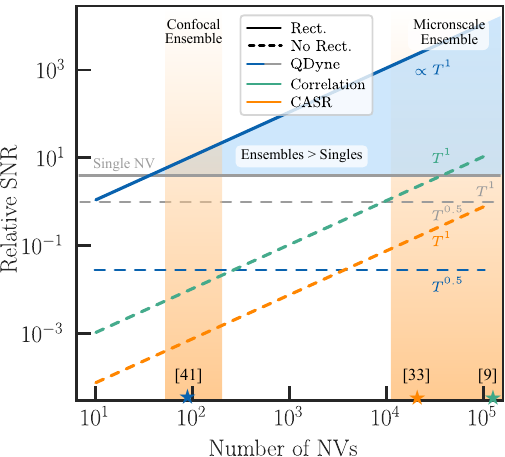}
    \caption{\label{fig:ensembles} \textbf{Sensitivity comparison of single and ensemble NVs after 5 min of measurement time at 2.7 T.}
    Rectified detection of statistical polarization (blue line) improves SNR in the PSD compared to single NV detection (grey line, blue area) requiring only $\sim 10^2$ NV centers.
    Non-rectified detection of statistical polarization (dashed blue line) cannot be improved by increasing the number of NV centers.
    Correlation protocols (dashed green line) as in \cite{aslam2017} can be used to detect statistical polarization with ensembles, but have unfavorable sampling rates, reducing the available number of scans.
    CASR detection of thermal polarization (dashed orange line) has favorable time scaling and can be applied to NV ensembles, but suffers from low thermal polarization.
    Orange areas indicate the two regions of NV NMR detection, using small ensembles in confocal spots and micronscale ensembles.
    The number of NVs utilized in recent ensemble NV NMR experiments \cite{Pagliero.25122024, Briegel.2025,liu2022surface} are indicated by stars.
    }
\end{figure}

Ensemble NVs are commonly used in NMR sensing due to their enhanced sensitivity compared to single NVs.
However, in the case of coherently averaged NMR sensing, only thermally (or hyperpolarized) AC signals have been detected so far using ensembles.
Even at elevated magnetic fields of \SI{2.7}{T}, the thermal polarization of proton spins only reaches $\sim 10^{-5}$, due to the low gyromagnetic ratio \cite{grafensteinCoherent2025}.
The statistical polarization in nanoscale NMR spectroscopy for shallow single NVs ($d < \SI{10}{nm}$) is almost 300 times larger than the thermal polarization at \SI{2.7}{T} \cite{grafensteinCoherent2025}.
While the statistical polarization of the target spins is still detectable in a classical ensemble measurement, the magnitude is significantly reduced, as the averaged readout of all NVs in the confocal spot also averages the effective statistical polarization of all individual NVs.
Generally, this effect results in a reduction of the detected signal $\propto N_\mathrm{NV}^{-1}$, while the noise is also reduced $\propto N_\mathrm{NV}^{-1}$, yielding no improvement with ensembles. 

By using the \textit{in situ} rectification protocol, this problem can be overcome, as the statistical polarization is rectified directly by each individual NV in the ensemble.
This way, each NV automatically produces a coherent signal, which is not averaged out over the ensemble.
This improvement is reduced by the infidelity of the rectification protocol through shot noise and charge state infidelities (see Supplementary Note 4).
Under realistic conditions, such as \SI{30}{\percent} charge state infidelity and an optimized shot-noise with $\alpha_\mathrm{opt} = 0.63 \pi$, the signal loss through rectification infidelity is only a factor of 3 compared to the perfectly deterministic case.
In cases, where $\alpha$ for each individual NV is distributed around an ensemble mean, the overall rectification fidelity is further reduced, while still allowing coherent averaging of the signal. 
As a result, the rectification protocols offer a significant advantage even for small ensembles in a confocal spot, while the advantage is much larger for bigger ensembles.

A comparison of the introduced rectification protocols and other state-of-art NV NMR protocols are shown in Fig. \ref{fig:ensembles} for a fixed measurement time of \SI{5}{min}.
The relative SNR of the power spectrum of each method for an ensemble is calculated for different numbers of NVs in the ensemble, assuming non-overlapping detection volumes and a linear increase of detected photons with number of NVs.
For details regarding the comparison of the calculated SNR, see Supplementary Note 7.
Recent studies on ensemble NV NMR spectroscopy have investigated two distinct operational regimes.
The first leverages a small ensemble of NV centers ($\sim 10^2$) confined within a diffraction-limited confocal laser spot, enabling nanoscale sensing. 
In contrast, the second employs a significantly larger ensemble of NVs ($\sim 10^4 - 10^5$) within a micronscale illumination area, facilitating measurements with enhanced signal strength and broader spatial coverage.
In the case of nanoscale NV ensembles, the SNR of rectified ensemble measurements closely matches that of the rectified single NV protocol, resulting in similar performance.
In this regime, the use of ensemble NVs does not necessarily lead to a higher SNR.
In contrast, for micronscale ensembles, the SNR enhancement relative to measurements with single NV centers can reach several orders of magnitude, highlighting a strong incentive to use ensemble NVs.
Notably, the introduced rectification protocol shows the highest SNR, even when accounting for the rectification infidelities arising from shot-noise and charge-state infidelities.
Incoherent averaging of statistical polarization is not compatible with ensemble measurements if no \textit{in situ} correlation protocol is used, as the statistical polarization is averaged out, yielding a constant low SNR.
While correlation-based protocols, which have so far been used for the detection of statistical polarization with single NVs, are generally compatible with ensemble NVs, they are less preferable as the acquisition of $m$ data points takes $2/(m+1)$ times longer compared to sequential measurement schemes.
This effectively reduces the SNR in the PSD by the same factor.
Coherent averaging protocols of thermal polarization via CASR protocols combine compatibility with NV ensembles and favorable SNR scaling with measurement time, but suffer from lower thermal polarization compared to statistical polarization, making it mostly suitable for large ensembles ($N_\mathrm{NV} > 10^5$).

\textit{Conclusion.---}In summary, we introduced and demonstrated two protocols to rectify statistical RF signals using a single NV center in diamond as quantum sensor, one based on the direct readout of the initial phase information (\textit{ex situ}) and one based on \textit{in situ} rectification.
These protocols enable efficient detection of statistical signals with high frequency resolution by improving the SNR scaling from $T^{0.5}$ to $T^{1}$.
These results are especially promising for nanoscale NMR experiments with NV ensembles, as the \textit{in situ} rectification potentially enables the efficient detection of the statistical polarization of the target spins with NV ensembles, combining high polarization from single NVs with high sensitivity from ensemble NVs.
While statistical polarization is a valuable resource at the nanoscale, its detection also introduces limitations on the coherence time of the target signal, as the target spins can diffuse out of the detection volume.
Maximizing the benefits of rectified statistical NMR requires increasing the target spin diffusion time, e.g., by immobilizing spins near the quantum sensor.
This was proposed theoretically \cite{Cohen.2020} and shown experimentally \cite{liu2022surface}.
By coherently detecting the statistical polarization, the sensitivity of nanoscale NMR spectroscopy could be increased by several orders of magnitude, potentially creating an important technique for the investigation of interfaces, thin-films or life sciences at the nanoscale.

In the final stages of preparing this manuscript, we learned of an independent but related work \cite{Spohn.2025}.

\textit{Acknowledgments.---}We acknowledge funding from the German Federal Ministry of Education and Research (BMBF) through the projects Clusters4Future QSens and DiaQnos.
We further acknowledge funding from the european union through project C-QuENS (Grant agreement no. 101135359), as well as the Carl-Zeiss-Stiftung via QPhoton Innovation Projects and the Center for Integrated Quantum Science and Technology (IQST).
R.M. acknowledges support from the International Max Planck Research School (IMPRS).
The authors would like to thank J. Hesselbarth and M. Lippoldt from the Stuttgart Institute for High Frequency Technologies for their valuable assistance with the VNA measurements of our microwave resonator. 

\bibliographystyle{naturemag}

\bibliography{references.bib}
\end{document}


\setcitestyle{numbers}
\maketitle

\maketitle

\section*{Supplementary Note 1: Theoretical Derivation of SNR Scaling}
In order to detect a signal of interest $s(t)$ on top of the noise $x(t)$, multiple traces are sequentially recorded $M_i(t_j) = s_i(t_j) + x_i(t_j)$, where we omit $t$ in the future for clarity. 
For this derivation we assume non-correlated white-noise with a standard deviation of $\sigma_x$.
When the phases of individual traces $i$ are not in sync between each new experimental run, e.g. in the case of the detection of statistically polarized nuclear spins in nanoscale NMR spectroscopy, on a large enough data set $\langle s_i\rangle \to 0$, as well as its Fourier representation $\langle\mathcal{F}(s)\rangle \to0$.
Hence, the autocorrelation $\langle M_i (t_j) M_i(t_j-\tau)\rangle$ or the auto power  $\langle|\mathcal{F}(M_i)|^2\rangle$ of the signal has to be averaged. 
In this case, $\mathcal{F}(M_i) = S_i + X_i$, where $S_i$ and $X_i$ are the respective Fourier representations of $s_i$ and $x_i$.
Calculation of the individual power spectra yields $|\mathcal{F}(M_i)|^2 = |S_i|^2 + |X_i|^2 + \mathrm{c.t.}$, where c.t. are the cross terms, which average to 0.
Still the standard deviation of $|X_i|^2$ is $\sigma_x^2$.
Averaging over $N$ experimental runs yields
\begin{align}
    \langle|\mathcal{F}(M_i)|^2 \rangle = \frac{1}{N}\sum_i^N |S_i|^2 + |X_i|^2 + \mathrm{c.t.} = |S|^2 + \langle|X|^2\rangle
\end{align}
Since the noise power $|X|^2$ is a random distributed (chi-squared distributed) quantity, the standard deviation of the average over $N$ repetitions scales with 
\begin{align}
    \sigma(\langle |X|^2\rangle) = \sigma_x^2/\sqrt{N}
\end{align}
the SNR improves with $\sqrt{N}$

In the case of a coherent signal with a predefined phase, direct averaging of the time traces becomes possible, as $\langle (s_i) \rangle \neq 0$.
Here, direct averaging yields 
\begin{align}
    \langle M_i \rangle =  \frac{1}{N}\sum_i^N  s_i + x_i = s + \langle x \rangle \thickspace ,
\end{align}
where again $x$ is a random-distributed quantity with $\sigma(\langle x_i \rangle ) = \sigma_x/\sqrt{N}$.
Calculation of the power spectrum yields $|\mathcal{F}(\langle M_i \rangle)|^2 = S^2 + |\langle X \rangle|^2$, where this time 
\begin{align}
    \sigma(|\langle X \rangle|^2) = \sigma(\langle x_i \rangle )^2 =\sigma_x^2/N \thickspace .
\end{align}
The noise scaling with number of measurements $N$ in the power-spectral density improves from $N^{-0.5}$ to $N^{-1}$ when coherent averaging is possible, compared to incoherent averaging.
Since the signal generally stays constant, the SNR scaling with $N$ also improves from $N^{0.5}$ to $N^1$.

\section*{Supplementary Note 2: Experimental Setup}

\begin{figure}[ht]
    \centering
    \includegraphics[width = 1 \textwidth]{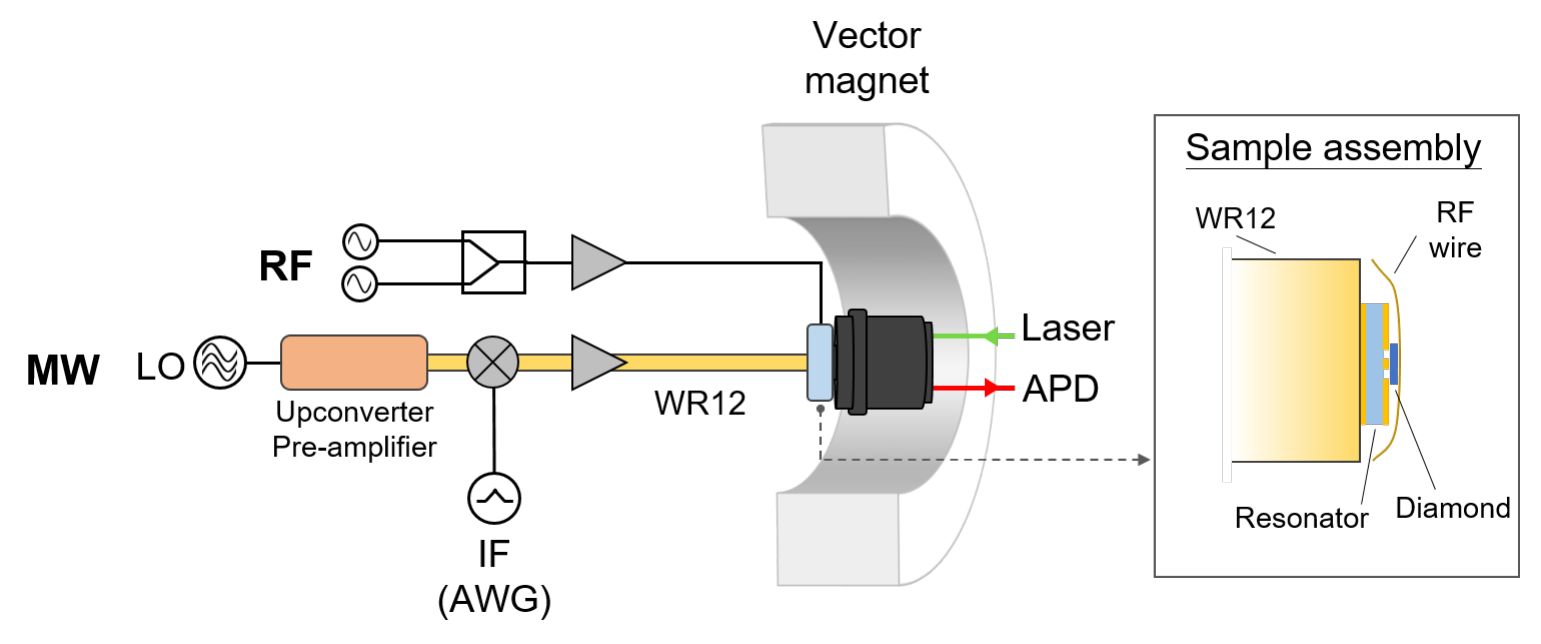}
    \caption{ \textbf{Experimental setup}.
    The setup is a confocal room temperature single NV experiment.
    The NV diamond is located in a superconducting magnet (\SI{2.7}{T}) and is addressed optically by a green laser through a high-NA objective.
    Red fluorescence is detected by an APD.
    The MW of the local oscillator (LO) is upconverted 6x and mixed with the pulsed MW of the AWG to generate \SI{73}{GHz} MW pulses.
    The RF nuclear spin control is combined with the weak artificial RF signal and delivered to the diamond via a copper wire on the back of the diamond.
    In the inset on the right the sample and resonator are shown in detail.
    } 
    \label{fig:Figure_S1_setup}
\end{figure}

A schematic representation of the setup is shown in Supplementary Fig. S\ref{fig:Figure_S1_setup}. 
A pulsed \SI{532}{nm} laser controlled via a TTL signal is used to optically excited the NV center.
The laser is focused onto the diamond with an oil objective with $\mathrm{NA} = 1.35$ and a working distance of \SI{300}{\mu m} (Olympus UPlanSApo 60$\times$). 
The red fluorescence is collected through the same objective, a \SI{50}{\mu m} pinhole and a \SI{650}{nm} long-pass filter by an avalanche photodiode (APD) for photon counting with a TimeTagger (Swabian Instruments). 
The diamond is glued on the resonator and the assembly is glued onto a rectangular waveguide (WR12). 
Additionally, a copper wire is placed on the backside of the diamond to provide RF for nuclear spin control and the target signal. 
Microwave pulses are created by mixing a continuous carrier frequency with a mixing pulse. 
To achieve this, a carrier frequency (Anritsu MG3697C) is upconverted 6 times in frequency, pre-amplified (S12MS, OML inc.) and mixed (SAGE-SFB-12-E2) with the pulsed output from the arbitrary waveform generator (Frequency \SIrange{100}{200}{MHz}, AWG, Keysight M9505A). 
The combined microwave is amplified (SAGE-SBP-7137633223-1212-E1 and SAGE-AMP-12-02540) and coupled to the resonator through a rectangular waveguide (WR12). 
The RF control of nitrogen spins is generated by the same AWG and the weak artificial RF signal ($f = \SI{166.666}{kHz}$) is sent by a signal generator (HP 33120A) with \SI{-43}{dBm} before amplification (AR150A250).

The (111) orientated diamond is double-sided polished and $^{12}$C enriched with a P1 concentration of about \SI{11}{ppb}. 
The thickness of the diamond is \SI{88}{\mu m}, so that we can see through it. 
The double-sided polished sapphire substrate with the crystal orientation $\alpha\text{-Al}_2\text{O}_3$ (0001) and the thickness of \SI{280}{\mu m} for the resonator is commercially available. 
The structure of the resonator is made by copper plating, photolithography, electroplating, reactive-ion-etching, and finally laser cutting.

\begin{table}[h]
    \centering
    \caption{Final parameters for the resonator. All the units are $\mu$m.}
    \label{tab1:res_params}
    \begin{tabular}{cccccccccc}
        \hline
        L$_{\text{res}}$  & W$_{\text{res}}$  & Gap & L$_{\text{p}}$ & W$_{\text{p}}$  & L$_{\text{tap}}$ & L$_{\text{w}}$  & W$_{\text{w}}$  & Sapphire size & THK$_\text{m}$ \\
        \hline
        470  & 100  & 45  & 667  & 1476 & 100  & 80  & 3  & $4428 \times 2668 \times 280$  & 3 \\
        \hline
    \end{tabular}
\end{table}

\begin{figure}[ht]
	\centering
	\includegraphics[width = 1 \textwidth]{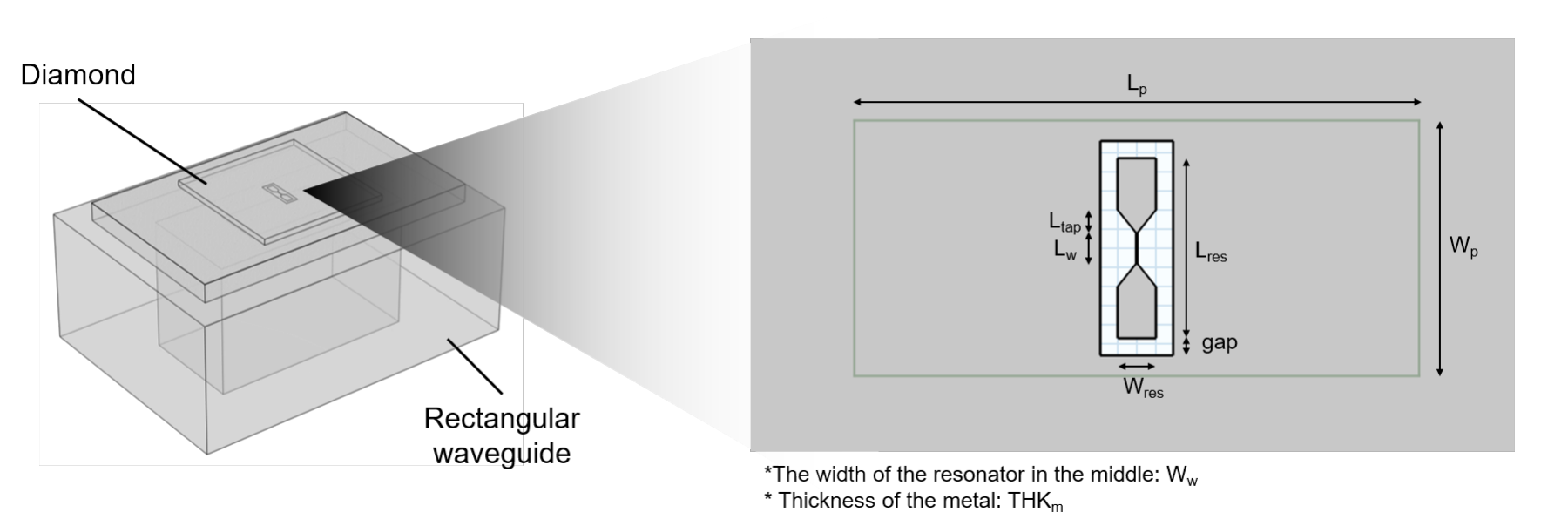}
	\caption{\textbf{Resonator geometry.}
	The 2D and 3D geometry of the microwave bow-tie resonator. 
    The rectangular waveguide and the diamond are included in the simulation.
    }
	\label{fig:Figure_S2_res}
\end{figure}

\section*{Supplementary Note 3: Optimization for the Microwave Resonator}
The 2D and 3D geometry of the resonator are shown in Supplementary Fig. S\ref{fig:Figure_S2_res}. The structure is modified from the literature \cite{aslamRes2015} and further optimized. The resonator is designed as a $\lambda /2$ coplanar waveguide (CPW) \cite{aslamRes2015} with a feeding patch on the backside facing the rectangular waveguide. The characteristic frequency is
$f_1 = \frac{c}{\lambda_0}\frac{1}{\sqrt{\epsilon_{\text{eff}}}}$, where $c$ is the speed of light, $\lambda_0$ is the wavelength of the microwave and $\epsilon_\mathrm{eff}$ is the effective permittivity.
The length of the resonator L$_\text{{res}}$ is $\lambda_0 /2$. 
Sapphire is chosen as the substrate to mitigate background fluorescence and to achieve a higher Q factor. Since $\epsilon_{\text{eff}}$ for sapphire is very different from the PCB used in the literature, we calculate the possible geometrical parameters and apply simulations afterwards. Numerical simulation are carried out by calculating the eigenfrequencies in COMSOL Multiphysics version 6.2. The eigenmodes and the microwave $B_1$ field parallel to the surface can be extracted. For simplicity, we use $B_1$ as the absolute magnetic field strength parallel to the surface at the position (x,y,z) = (0,0,10) $\mu$m above the resonator in the following text.
$\epsilon_{\text{eff}}$ is analytically solved by calculating the elliptical functions iteratively based on the conductor backed coplanar waveguide (CBCPW) model to match an impedance of \SI{50}{\ohm}. W$_\text{res}$ and L$_\text{res}$ are obtained with the initial guess of gap. All the other values are taken from the literature \cite{aslamRes2015} as an initial guess. We sweep L$_\text{res}$, W$_\text{res}$, and the gap to find the structure with desired frequency. The thickness of the metal is assumed to be zero to simplify the simulation. Notably, it is mentioned that the thickness of the metal does not affect the impedance much but affects $\epsilon_{\text{eff}}$, thus leading to the deviation of the resonating frequency \cite{simonsCoplanar2001}. For this reason, the optimization target is set to find the maximum $B_1$ in the range of 75 $\pm$ 3\,GHz.

The optimized results are shown in Supplementary Fig. S\ref{fig:Figure_S3}a. The best six cases are highlighted with star markers. We then include the thickness of metal THK$_\text{m}$ = 3\,$\mu$m in the simulation. The final parameters are shown in Supplementary Tab. S\ref{tab1:res_params}. L$_{\text{tap}}$ is also swept and other types of taper such as exponential shape and Klopfenstein taper \cite{klopfensteinTransmission1956} are also considered and simulated, but no obvious improvements are observed. The resonator is finally fabricated by ourselves.

Supplementary Fig. S\ref{fig:Figure_S4} shows single NVs and the resonator with little background fluorescence. The Rabi frequency at different resonant frequencies is measured to evaluate the performance of the resonator. The Q factor obtained form the Rabi frequency drops to 42 compared to the result measured by VNA. This difference could be explained by the environment of measurement, the coupling efficiency between the rectangular waveguide and the resonator, and the radiation loss when driving NVs. Further improvements can still be done by making the critical coupling condition and tuning the radiation map.

\begin{figure}[ht]
	\centering
	\includegraphics[width = 1 \textwidth]{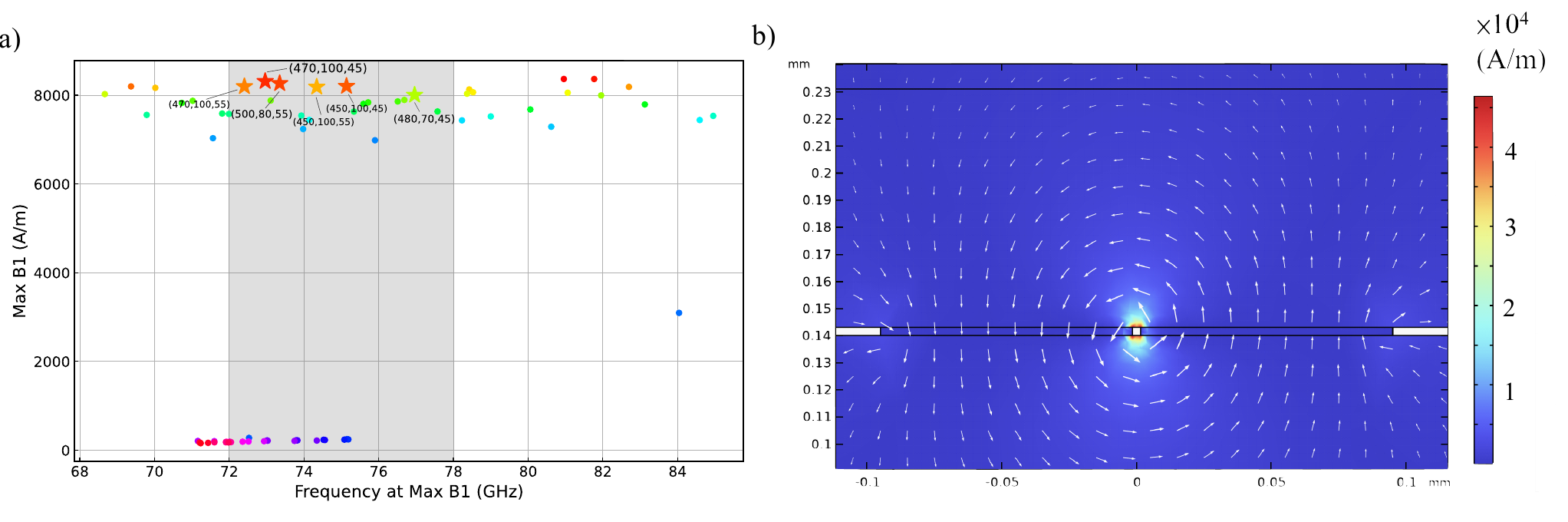}
	\caption{\textbf{Parameter sweeping for the resonator and microwave $B_1$ field}.
	\textbf{a)} The parameter sweeping for the resonator.
    The six best results in the desired frequency range are highlighted as star markers. 
    \textbf{b)} The microwave $B_1$ field projected on the XZ cut plane in the eigenmode. The color stands for the $B_1$ field strength and the white arrows represent direction of the field.
    }
	\label{fig:Figure_S3}
\end{figure}
	
\begin{figure}[ht]
	\centering
	\includegraphics[width = 1 \textwidth]{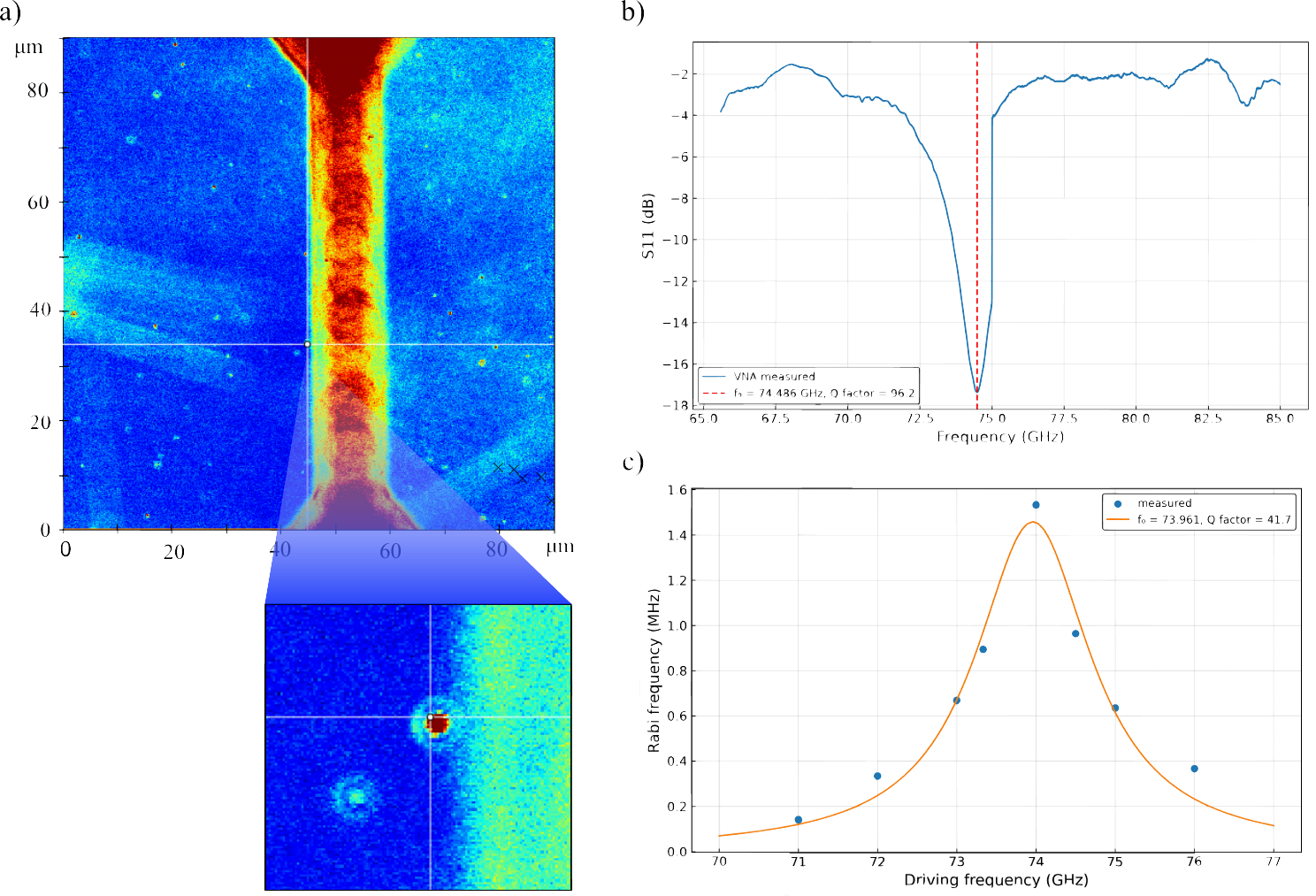}
	\caption{\textbf{Confocal map and the performance of the resonator}.
    \textbf{a)} The confocal map of 90 $\times$ 90\,$\mu$m.
    The inlet shows two single NVs next to the resonator.
    \textbf{b)} The performance of the resonator measured by VNA ranging 65 to 85 GHz.
    The red line is the resonating frequency.
    The Q factor is obtained by the conventional 3 dB method.
    \textbf{c)} The relation of the Rabi frequency and the operating microwave frequency.
    The Q factor is obtained by the Lorentzian fitting (orange curve) of the experimental data (blue dots).
    }
	\label{fig:Figure_S4}
\end{figure}

\begin{figure}[ht]
	\centering
	\includegraphics[width = 1 \textwidth]{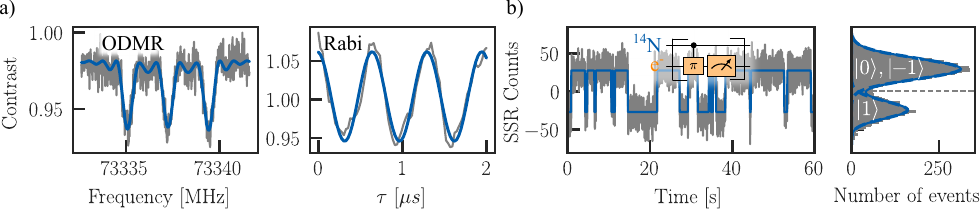}
	\caption{\textbf{Spin manipulation at 3T.}.
    \textbf{a)} Optically detected magnetic resonance (ODMR) of the electron spin at \SI{73.37}{GHz} with $^{14}$N hyperfine splitting and Rabi oscillations with frequency $\Omega_R = \SI{1.5}{MHz}$.
    \textbf{b)} Single-Shot readout (SSR) of the nitrogen spin state. Jumps in the fluorescence time trace show real time spin transitions of the nitrogen spin. 
    }
	\label{fig:Figure_3T_ODMR}
\end{figure}

In this implementation, the detection of optically detected magnetic resonance (ODMR) of the electron spin is possible at \SI{73}{GHz} or \SI{2.7}{T} (Supplementary Fig. S\ref{fig:Figure_3T_ODMR}a).
The hyperfine coupling to the adjacent $^{14}$N is clearly visible in the three-fold splitting of the ODMR peak, allowing easy access to the quantum memory.
Rabi oscillations of the electron spin are visible up to a Rabi frequency $\Omega_R$ of \SI{1.5}{MHz}, showing an improvement compared to previous resonator designs \cite{aslamRes2015} and allowing robust spin control for sensing protocols.
The lifetime of the quantum memory, under perturbations induced by sensor spin manipulation is obtained via sequential quantum non-demolition measurements, where the change of the memory state can be tracked in real-time (Supplementary Fig. S\ref{fig:Figure_3T_ODMR}b) \cite{Neumann.2010}.

\section*{Supplementary Note 4: Rectification Fidelity}
\begin{figure}[ht]
\centering
\includegraphics[width = \textwidth]{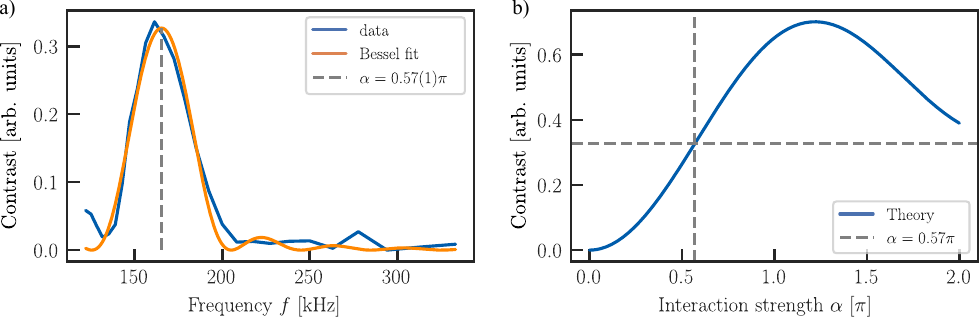}
\caption{ \textbf{DD Noise Spectroscopy}. 
\textbf{a)} RF detection using XY8-1 noise spectroscopy.
The resonance peak at $f = \SI{166}{kHz}$ shows an interaction strength of $\alpha = \SI{0.57(1)}{\pi}$, indicating a signal amplitude of $B_\mathrm{signal} = \SI{664(15)}{nT}$.
The signal is fitted with a 0th-order Bessel function of the first kind $J_0$, modulated with a sinc.
\textbf{b)} Determination of the interaction strength $\alpha = 0.57 \pi$. 
\label{fig:DD_Detection}}
\end{figure}

\begin{figure}[ht]
    \centering
    \includegraphics[width = \textwidth]{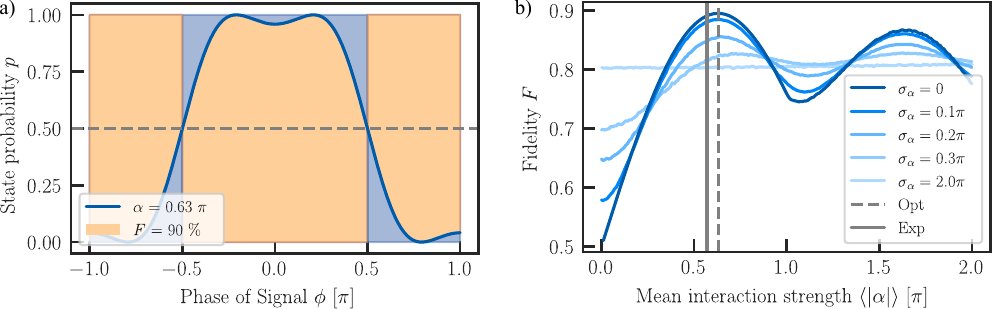}
    \caption{ \textbf{Shot noise limit of the rectification}. 
    \textbf{a)} Probability $p_0$ of the $^{14}$N memory to be in state $\ket{0}$ based on the signal phase $\phi$ for an interaction strength $\alpha = 0.63 \pi$.
    The blue shaded area represents ideal binary decision-making, where $p_0 = 1$ for $\phi = [\SI{90}{\degree},\SI{270}{\degree}]$ and 0 else.
    The real, shot noise limited rectification fidelity $F$ is given by the mean orange area.
    \textbf{b)} Shot-noise limited rectification fidelity $F$ with respect to the interaction strength $\alpha$. 
    A maximum fidelity of $\SI{90}{\percent}$ is achieved for $\alpha = 0.63 \pi$.
    The experimental parameter $\alpha = 0.57 \pi$ corresponds to a fidelity of $\SI{89}{\percent}$.
    In the case of an ensemble distribution of $\alpha$, where $\sigma_\alpha \neq 0$ the maximum rectification fidelity is slightly reduced.
    \label{fig:Fidelity}}
\end{figure}

\begin{figure}[ht]
    \centering
    \includegraphics[width = \textwidth]{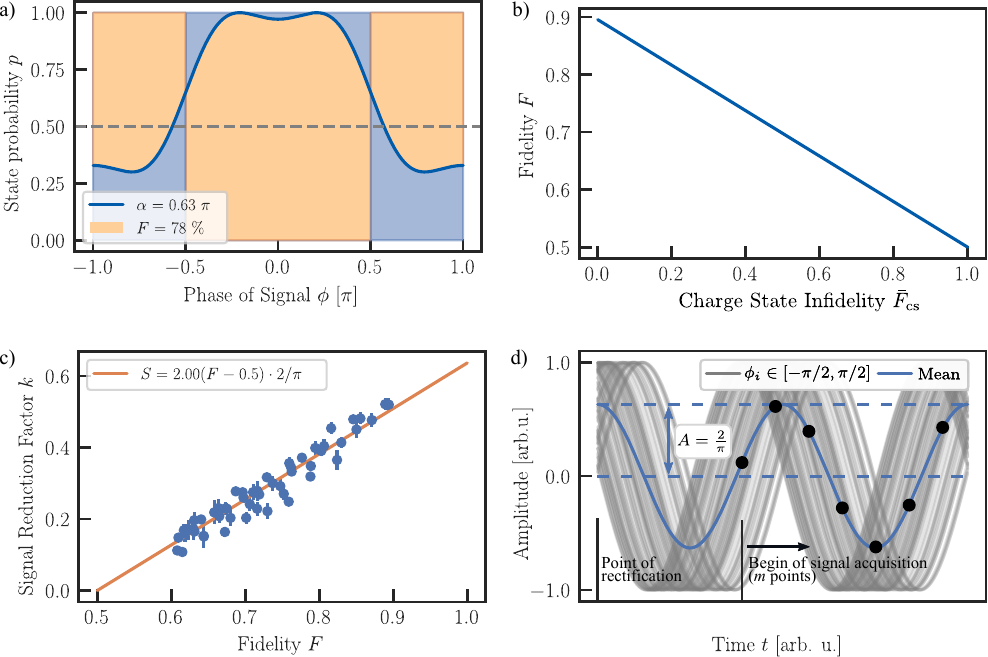}
    \caption{ \textbf{Charge state infidelity}. 
    \textbf{a)} Probability $p_0$ of the $^{14}$N memory to be in state $\ket{0}$ based on the signal phase $\phi$ for an interaction strength $\alpha = 0.63 \pi$ when charge state infidelities $\bar{F}_{cs}$ are present during the sensing time.
    The probability $p_0$ is shifted according to $\bar{F}_{cs}$.
    \textbf{b)} Influence of the charge state infidelity $\bar{F}_{cs}$ on the rectification fidelity $F$ for $\alpha = 0.63 \pi$. 
    The charge state infidelity linearly reduces the rectification fidelity $F$.
    \textbf{c)} Signal reduction factor $k$ is linearly reduced with decreasing fidelity $F$ with a maximum value of $\frac{2}{\pi}$.
    The blue points are the signal strengths obtained from rectified QDyne simulations with different values of $\alpha$ and $\bar{F}_\mathrm{cs}$. 
    \textbf{d)} Schematic of averaging rectified oscillating signals with random initial phases $\phi_i$.
    Rectification takes place at $t = 0$ such that $\phi_i \in [-\pi/2, \pi/2]$ resulting in a reduction of the signal amplitude by $\frac{2}{\pi}$.
    \label{fig:Fidelity_cs}
    }
\end{figure}

\subsubsection*{DD noise detection}
In a DD-noise detection experiment the spacing $\tau$ between the $N_p$ $\pi$ pulses is swept to find the frequency of a target signal $f_\mathrm{t}$. 
During one DD sensing block, the NV center acquires a phase of 
\begin{align}
    \theta = \alpha \cos{(\phi)} \mathrm{sinc}(N_p \pi \delta \tau) \thickspace ,
\end{align}
where $\alpha$ is the interaction strength between the target signal and the electron spin, $\phi$ is the initial phase of the target signal and $\delta = \frac{1}{2\tau} - f_\mathrm{t}$.
The interaction strength $\alpha$ is related to the sensing time $\tau_\mathrm{sens} = \SI{24}{\mu s}$, the applied signal amplitude $B_\mathrm{signal}$ and the gyromagnetic ratio of the electron spin $\gamma_e = \SI{28040}{MHz \per T}$ via \cite{Degen.2017}
\begin{align}
    \alpha = 2 \pi (\frac{2}{\pi} B_\mathrm{signal} \gamma_e \tau_\mathrm{sens}) \thickspace .
\end{align}
After the second $\pi$/2 pulse the probability to find the electron in $\ket{0}$ is given by 
\begin{align}
    p_0 = \frac{\cos(\alpha \cos(\phi) \mathrm{sinc}(N_p \pi \delta \tau))+1}{2} \thickspace .
\end{align}
For randomized initial phases, the overall signal is given by:
\begin{align}
	S &= 1-\left \langle \frac{\cos(\alpha \cos(\phi) \mathrm{sinc}(N_p \pi \delta \tau))+1}{2} \right \rangle \\
	&= 0.5-\frac{1}{2\pi}\int_0^{\pi} \cos(\alpha \cos(\phi) \mathrm{sinc}(N_p \pi \delta \tau)) \mathrm{d}\phi\\
    &= \frac{1-J_0(\alpha \mathrm{sinc}(N_p \pi \delta \tau))}{2} \thickspace , \label{eq:alpha}
\end{align}
where $J_0$ is the 0th order Bessel function of the first kind.
By fitting the experimental DD-detection peak (Supplementary Fig. \ref{fig:DD_Detection} a), the interaction strength $\alpha$ can be obtained (Supplementary Fig. \ref{fig:DD_Detection} b).
For the XY8-1 detection sensing block, we obtain $\alpha = 0.57 \pi$.
This interaction strength is comparable to standard NV-NMR experiments, as shown in Supplementary Tab. \ref{tab:alpha_param}.

\begin{table}[]
\centering
\caption{Typical interaction strength $\alpha$ in NV-NMR experiments. Values are calculated from the signal contrast according to Eq. \ref{eq:alpha}.}
\label{tab:alpha_param}
\begin{tabular}{cccccccc}
    \hline
    NV-type & Depth [nm] & Target & $\tau_\mathrm{sens}$ [$\mu$s] & $\alpha$ [$\pi$] & Ref. \\ \hline
    Single & 5 - 10 & Oil ($^1$H) & 5 - 50 & 0.1 - 0.5 & \cite{DeVience.2015} \\ 
    Single & 7 & hBN (single layer, $^{10}$B, $^{11}$B) & 500 & 0.3 - 0.4 & \cite{11BLovchinsky.2017} \\ 
    Single & 3 - 7 & Oil ($^1$H) & 10 - 30 & 0.5 - 0.8 & \cite{Fukuda.2018} \\ 
    Ensemble & 3 - 10 & Oil ($^1$H) & 18 - 180 & 0.6 & \cite{Fuhrmann.2024} \\ 
    & & & & & \\
    Single & $\sim$ 1000 & RF & 24 & 0.57 & This work \\ \hline
\end{tabular}
\end{table}

\subsubsection*{Imperfection of the Rectification}
To obtain a phase sensitive sensor response, the second $\pi/2$ pulse in the Qdyne sensing block, as well as the phase rectification block, is in $y$ direction compared to $x$ direction in common DD-detection protocols.
After the second $\pi/2$ pulse the probability to find the electron in $\ket{0}$ is given by 
\begin{align}
    p_0 = \frac{\sin(\alpha \cos(\phi))+1}{2} \thickspace .
\end{align}
The information about $\phi$ stored in $p_0$ is then transferred from the electron spin to the nitrogen memory spin and is used for the rectification through readout or \textit{in situ}.
For an ideal, deterministic rectification process, $p_0$ would be 1 for $\phi \in \left[-\pi/2, \pi/2 \right]$ and 0 everywhere else, producing a rectification fidelity $F$ of 1 (see Supplementary Fig. \ref{fig:Fidelity} a).
In contrast, the shot-noise limited fidelity $F_\mathrm{SN}$ due to probabilistic state projections is described by
\begin{align}
    F_\mathrm{SN} &= \frac{1}{\pi} \int_{-\frac{\pi}{2}}^{\frac{\pi}{2}} p_0 \mathrm{d} \phi\\
    &= \frac{1}{\pi} \int_{-\frac{\pi}{2}}^{\frac{\pi}{2}} \frac{\sin(\alpha \cos(\phi))+1}{2} \mathrm{d} \phi
\end{align}
Numerical integration of this expression for different $\alpha$ yields a maximum fidelity of $F_\mathrm{SN} \approx \SI{90}{\%}$ at the optimal $\alpha_\mathrm{opt} = 0.63 \pi$.
At our experimental parameter $\alpha_\mathrm{exp} = 0.57 \pi$ the shot-noise limited rectification fidelity is \SI{89}{\percent}.
If alpha is not constant, but is instead represented by a gaussian distribution with a mean of $\langle |\alpha| \rangle$ and a standard deviation of $\sigma_\alpha$ the average rectification fidelity is affected slightly (Supplementary Fig. \ref{fig:Fidelity} b).
This can be the case for measurements of freely diffusing target molecules or measurements with an ensemble of NV centers, where each NV center might have a different depth $d$.

A second source of infidelity is the charge state infidelity $\bar{F}_\mathrm{cs}$ of the NV center after laser illumination.
With a probability of $\approx \SI{30}{\%}$ the NV center is not in the negative NV$^-$ charge state, but the neutral NV$^0$ state.
In this state the NV center is insensitive to the target signal, thus the memory state stays in $\ket{0}$, independent of the actual phase of the signal.
As a result
\begin{align}
    \bar{p}_0 &= ( 1 - \bar{F}_\mathrm{cs}) p_0 + \bar{F}_\mathrm{cs}\\
    &=( 1 - \bar{F}_\mathrm{cs}) \frac{\sin(\alpha \cos(\phi))+1}{2} + \bar{F}_\mathrm{cs}
\end{align}
Therefore, the probabilities to be in state 0 or 1 ($p_0$ and $p_1 = 1 - p_0$) are no longer symmetric and the overall fidelity $F$ is defined via
\begin{align}
    F &= \frac{1}{2\pi} \left(\int_{-\frac{\pi}{2}}^{\frac{\pi}{2}} \bar{p}_0  \, \mathrm{d} \phi + \int_{\frac{\pi}{2}}^{\frac{3\pi}{2}} \bar{p}_1  \, \mathrm{d} \phi \right)\\
    &= \frac{1}{2\pi} \left( \int_{-\frac{\pi}{2}}^{\frac{\pi}{2}} \left[(1-\bar{F}_\mathrm{cs})p_0 + \bar{F}_\mathrm{cs}  \right]\, \mathrm{d} \phi + \int_{\frac{\pi}{2}}^{\frac{3\pi}{2}} (1-\bar{F}_\mathrm{cs}) p_1  \, \mathrm{d} \phi  \right)\\
    &= \frac{\bar{F}_\mathrm{cs}}{2} + (1-\bar{F}_\mathrm{cs}) \frac{1}{2\pi}\left(\int_{-\frac{\pi}{2}}^{\frac{\pi}{2}} p_0  \, \mathrm{d} \phi + \int_{\frac{\pi}{2}}^{\frac{3\pi}{2}} p_1  \, \mathrm{d} \phi \right) \\
    &= \frac{\bar{F}_\mathrm{cs}}{2} + (1-\bar{F}_\mathrm{cs}) F_\mathrm{SN} \thickspace .
\end{align}
The linear dependence on $\bar{F}_\mathrm{cs}$ is visualized in Supplementary Fig. \ref{fig:Fidelity_cs}b.
Thus, in the case of $\alpha_\mathrm{opt} = 0.63 \pi$ and a charge state infidelity of \SI{30}{\percent} the overall rectification fidelity reduces from \SI{90}{\percent} to \SI{78}{\percent}.
For a binary signal, that can only be fully constructive or destructive, the influence of the overall fidelity of $F$ on the normalized signal $S$ is given by the reduction factor
\begin{align}
    k &= F - (1 - F) = 2F - 1 \thickspace .   
\end{align}
Due to the continuous random phase $\phi$ of the oscillating traces, not all time traces have the same contribution to the overall signal, as not all of them will be fully constructive or destructive.
Thus, the influence of $F$ on $S$ depends on the interaction strength $\alpha$.
To investigate this effect, $S$ has been numerically simulated for different parameters of $\alpha$ and $\bar{F}_\mathrm{cs}$.
The resulting relationship is shown in Supplementary Fig. \ref{fig:Fidelity_cs}c. 
This reproduced the idealized relationship and confirmed the theoretical approach.
In the experimental case of $F = \SI{78}{\percent}$, this leads to a signal loss in the PSD of $1-k^2 \approx \SI{68}{\percent}$ ($k = 0.57$) compared to the deterministic case without charge state infidelity.
However, it has to be kept in mind, that even in the fully deterministic rectification case, the amplitude of an averaged signal is still reduced, because the oscillating signals are not averaged fully coherent, but instead only over the space of $\phi_i \in [-\pi/2, \pi/2]$, resulting in an additional signal reduction of
\begin{align}
    \frac{1}{\pi} \int_{-\pi/2}^{\pi/2} \cos(\phi) d\phi = \frac{2}{\pi} \thickspace .
\end{align}
This means, that even in the case of perfect rectification fidelity, without shot-noise and charge state-infidelity, at most $2/\pi$ of the signal amplitude can be restored with a binary rectification process of an oscillating signal with a continuous phase distribution (Supplementary Fig. \ref{fig:Fidelity_cs} d).
As a result, the final signal reduction factor $k$ is given by 
\begin{align}
    k = \frac{2}{\pi}(2F -1) \thickspace . \label{eq:red_factor}
\end{align}
With an experimental fidelity of $F \approx \SI{78}{\percent}$, the achievable signal in the PSD is reduced by $\SI{87}{\percent}$ ($k = 0.36$).

\section*{Supplementary Note 5: Experimental Details}
In the rectification experiments, the nitrogen spin and the electron spin are initialized at the beginning of the sequence.
The nitrogen spin is initialized into $\ket{1}$ through an effective SWAP gate consisting of an electron $\pi$ rotation conditioned on the nitrogen state $\ket{0}$, followed by a nitrogen $\pi$ conditioned on the electron $\ket{-1}$ state and re-initialization of the electron spin by a laser pulse.
The spin state of the nitrogen spin is confirmed via single-shot readout (SSR) according to standard NV experimental procedures (see e.g. Ref. \cite{Neumann.2010, Waldherr.2014}).
Experimental runs where the nitrogen spin was wrongly initialized were discarded.
Population in the nitrogen $|-1\rangle$ state were not initialized into $|1\rangle$ but could be included for an optimization of the experiment.
In total, we achieved an initialization probability of $\approx \SI{60}{\percent}$, meaning that in the rectification protocols \SI{40}{\percent} of the $N = 25000$ experimental traces were discarded.
In our experimental implementation both SSR blocks are repeated 3000 times ($\approx$ \SI{12}{ms}) and the sequential measurement (\textit{Qdyne}) block is repeated 4000 times ($\approx$ \SI{110}{ms}).
In this work, the measurement time is limited to \SI{110}{ms} to mimic realistic conditions with immobilized target spins, but it could easily be extended up to the memory lifetime of $\SI{1.9}{s}$, as the artificial target signal has no inherent lifetime.
The whole experiment is repeated $N$ times to improve the SNR of the PSD and the initial phase of the target signal is randomized for each repetition by adding random delay times to the end of the sequence after each repetition.
The sensing block consists of an XY8-1 decoupling sequence with a resonant spacing of $\tau = \SI{3}{\mu s}$, resulting in a total sensing time of $\tau_\mathrm{sens} = \SI{24}{\mu s}$.
A summary of all repetition numbers $n$, sequence duration $t_\mathrm{seq}$, total measurements times $t_\mathrm{meas}$ with and without experimental overhead through data transfer, optical and ODMR refocus, are given in Supplementary Tab. \ref{tab:exp_param}.
The total measurement time can be further optimized, as we ran refocus sequences every 1000 averages, while the stability of our system would allow for significantly less refocussing.
In the case of the \textit{ex situ} rectification, the time traces have been added or subtracted, based on the result of the second SSR.
In the standard QDyne and \textit{in situ} rectification no subtraction is necessary.
In all experiments, the PSD has been calculated through $|\mathrm{FFT}|^2$ (before or after averaging of the traces), and the SNR was calculated from the obtained PSD as the difference between the power at the resonance frequency $P_\mathrm{signal}$ and the power of the noise floor $P_\mathrm{noise}$, over the noise as the root-mean-square (RMS) of the noise floor $\sigma_\mathrm{P_{noise}}$ according to common practice in the NMR community.

\subsubsection*{Details of the experimental sequence}
In the case of both rectification protocols, the electron and nitrogen spin are initialized into $|0\rangle$ and $|1\rangle$, respectively, forming the state $|\Psi\rangle = |0\rangle |1\rangle$.
Next, the electron spin is used to sense the initial phase $\phi_i, \thickspace i = 0, ..., N-1$ of the target signal, which transforms the overall state into
\begin{align}
    |\Psi \rangle &= (a_i|0\rangle + b_i |1\rangle)|1\rangle \\
    a_i &= \sin\left(\alpha \cos(\phi_i) + \frac{\pi}{4}\right) \thickspace , \thickspace b_i = \cos\left(\alpha \cos(\phi_i) + \frac{\pi}{4}\right) \thickspace ,
\end{align}
where the phase information of $\phi_i$ is stored in the coefficients $a_i$ and $b_i$.
Applying a controlled $\pi$-pulse on the nitrogen spin (blue gate in Fig. 3 a of the main text) before re-initializing the electron spin by applying a green laser pulse, transfers the phase information of $\phi_i$ onto the population of the nitrogen memory spin:
\begin{align}
    |\Psi \rangle &= |0\rangle (a_{i}|0\rangle + b_{i} |1\rangle) \thickspace .
\end{align}
Now the electron spin can be used to directly read out the spin state of the nitrogen state via single-shot readout \cite{Neumann.2010} (\textit{ex situ} protocol), or it can be used to sequentially probe the phase of the target signal (\textit{in situ} protocol).

\textbf{\textit{ex situ} protocol:}
After reading out the nitrogen spin classically and obtaining $0$ or $1$, the nitrogen spin is no longer used for the protocol.
The electron spin sequentially probes the instantaneous phase $\phi_{i,j}, \thickspace j = 0, ..., m-1$ of the target signal $m-$times and is read out each time.
The spin state of the electron spin before each readout is given by
\begin{align}
    |\Psi \rangle_e &= (a_{i,j}|0\rangle + b_{i,j} |1\rangle) \\
    a_{i,j} &= \sin\left(\alpha \cos(\phi_{i,j)} + \frac{\pi}{4}\right) \thickspace , \thickspace b_{i,j} = \cos\left(\alpha \cos(\phi_{i,j}) + \frac{\pi}{4}\right) \thickspace .
\end{align}
As the phase $\phi_{i,j}$ oscillates during the $m$ readouts, the oscillating time trace $M_i$ is acquired.
By multiplying each obtained time trace with +1 or -1 based on the readout of the memory qubit ($0$ or $1$), the time traces can be coherently averaged.
Imperfections of the rectification are discussed in Supplementary Note 4.

\textbf{\textit{in situ} protocol:}
Similarly, in the \textit{in situ} protocol the spin states before the final controlled $\pi$-rotation are given by
\begin{align}
    |\Psi \rangle &= (a_{i,j}|0\rangle + b_{i,j} |1\rangle)(a_{i}|0\rangle + b_{i} |1\rangle) \thickspace ,
\end{align}
where the information about the initial phase $\phi_i$ is stored in the population basis of the nitrogen spin and the information about the current phase is stored in the population basis of the electron spin.
Applying the final controlled $\pi$-gate in the form of 
\begin{align}
    g = (|1\rangle \langle0| + |0\rangle \langle1|)|1\rangle \langle1| + (|1\rangle \langle1| + |0\rangle \langle0|)|0\rangle \langle0|
\end{align}
results in the spin state
\begin{align}
    |\Psi \rangle &= a_{i}(a_{i,j}|0\rangle + b_{i,j} |1\rangle)|0\rangle + b_{i} (b_{i,j}|0\rangle + a_{i,j} |1\rangle)|1\rangle \thickspace,
\end{align}
clearly showing that the obtained time trace obtained through the optical readout of the electron spin is directly \textit{in situ} inverted, based on the coefficients $a_{\textit{i}}$ and $b_{\textit{i}}$.
Again, the same rectification infidelities as in the \textit{ex situ} protocol are present, which are discussed in Supplementary Note 4.

\textbf{\textit{Qdyne} protocol (no rectification):}
For the Qdyne protocol, no memory qubit is used, as the full experimental sequence is given only by the sequential measurement block, as shown in the \textit{ex situ} protocol in Fig. 3 a of the main text.
After each sensing step the electron spin is in the state
\begin{align}
    |\Psi \rangle_e &= (a_{i,j}|0\rangle + b_{i,j} |1\rangle) \thickspace ,
\end{align}
which is directly read out with a laser pulse.
As no rectification is applied, direct averaging of the time traces $M_i$ would lead to zero signal.
Therefore, the PSD of each time trace is averaged instead.

\begin{table}[]
\centering
\caption{Experimental parameters of the experiments.}
\label{tab:exp_param}
\begin{tabular}{cccccccc}
    \hline
    Measurement & $n_\mathrm{SSR_1}$ & $n_\mathrm{SSR_2}$ & $m$ (Qdyne rep.) & $t_\mathrm{seq}$ [ms] & $N$ & $t_\mathrm{meas, raw}$ [min] & $t_\mathrm{meas, total}$ [min]\\ \hline
    QDyne & - & - & 4000 & 110 & 25000 & 46 & 103\\
    Rect. \textit{ex situ} & 3000 & 3000 & 4000 & 137 & 14869 & 57 & 113\\
    Rect. \textit{in situ} & 3000 & - & 4000 & 118 & 14422 & 49 & 100\\ \hline
\end{tabular}
\end{table}

\section*{Supplementary Note 6: Data Analysis}
\begin{figure}[ht]
    \centering
    \includegraphics[width = \textwidth]{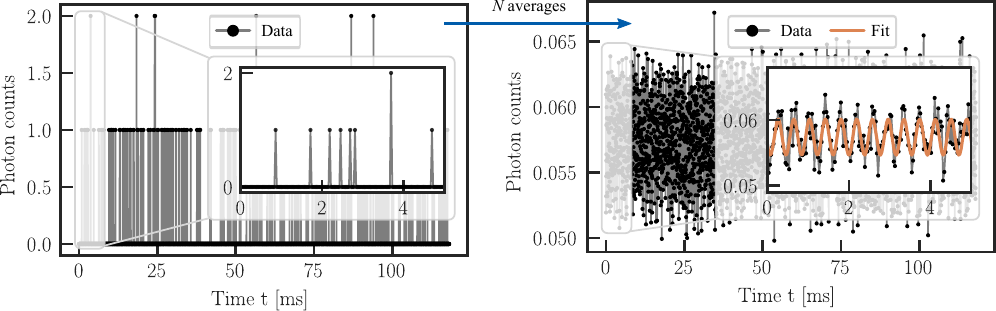}
    \caption{ \textbf{Time trace averaging}. 
    \textbf{a)} A single detected photon trace $M_i$ consisting of $m=4000$ points.
    Due to the low number of average photons detected per readout, the photon-shot-noise dominates.
    \textbf{b)} Time trace after coherently averaging $N = 14400$ traces.
    The photon-shot-noise is averaged out and the underlying oscillating signal $S$ (orange fit) becomes visible.
    Without successful rectification the signal would also average out, resulting in a flat line.
    This is the underlying data of the final data point of the \textit{in situ} experiment in Fig. 3 c in the main text.
    The SNR is obtained from the PSD after taking $|$FFT$|^2$ of the averaged time trace.
    \label{fig:trace_averaging}}
\end{figure}

The data analysis and SNR discussion is presented here in detail for the example of the \textit{in situ} experimental data from Fig. 3 c in the main text.
For each of $N$ averages a photon time trace with $m$ points is recorded, in which the underlying oscillating signal
\begin{align}
    s_i(t) = \bar{n}_\mathrm{ph} (1 + \frac{c}{2} \cos(f_0 t + \phi_i))
\end{align}
is hidden under the photon shot noise (Supplementary Fig. \ref{fig:trace_averaging} a).
Here, $\bar{n}_\mathrm{ph}$ denotes the average number of detected photons, $c$ is the optical contrast of the NV center and $\phi$ is the random initial phase of the signal.
The detected frequency $f_0$ of the signal may differ from the applied frequency $f$ due to undersampling, but $f$ can be easily recalculated based on the sampling frequency.
Due to the low number of photons per readout ($\bar{n}_\mathrm{ph} = 0.057$), the photon shot noise constitutes the main noise source, while quantum projection noise and readout noise remain largely insignificant.
Coherent averaging of the time traces based on the rectification protocol reduces the photon shot noise (Supplementary Fig. \ref{fig:trace_averaging} b).
The amplitude of the oscillating signal in the averaged time trace is given by
\begin{align}
    S = \frac{\bar{n}_\mathrm{ph} c}{2} k ,
\end{align}
where $k$ is the reduction factor introduced by the infidelity of the rectification process (Eq. \ref{eq:red_factor}).
In the power spectrum of the discrete time trace with $m$ points this yields the signal power at the frequency $f_0$
\begin{align}
    P_\mathrm{signal} = \left( \frac{m}{2}S\right)^2 = \left(\frac{1}{2}\frac{\bar{n}_\mathrm{ph} c m}{2} k\right)^2
\end{align}
in a single frequency bin.

Assuming uncorrelated photon-shot-noise (\textit{white noise}), the power spectrum of a single time trace is flat with a baseline of 
\begin{align}
    P_\mathrm{noise, 1} = \sqrt{\bar{n}_\mathrm{ph} m}^2 = \bar{n}_\mathrm{ph} m
\end{align}
and a variance of the baseline of 
\begin{align}
    \sigma_{P_\mathrm{noise, 1}} = P_\mathrm{noise, 1} = \bar{n}_\mathrm{ph} m \thickspace .
\end{align}
As described in Supplementary Note 1, coherent (incoherent) averaging of $N$ traces reduces both $P_\mathrm{noise}$ and $\sigma_{P_\mathrm{noise}}$ by a factor of $1/N$ $(1/\sqrt{N})$, resulting in 
\begin{align}
    P_\mathrm{noise,N}^\mathrm{coh} &= \sigma_{P_\mathrm{noise, N}^\mathrm{coh}}  = \frac{\bar{n}_\mathrm{ph} m}{N} \thickspace , \\ 
    P_\mathrm{noise,N}^\mathrm{incoh} &= \sigma_{P_\mathrm{noise, N}^\mathrm{incoh}}  = \frac{\bar{n}_\mathrm{ph} m}{\sqrt{N}} \thickspace .
\end{align}

Calculating the signal-to-noise ratio as the difference between the signal power and baseline power over the standard deviation of the baseline yields
\begin{align}
    \mathrm{SNR_{coh}} &= \frac{P_\mathrm{signal} - P_\mathrm{noise}}{\sigma_{P_\mathrm{noise}}} = \frac{\bar{n}_\mathrm{ph} m c^2 k^2}{16} N  -1 \approx \frac{\bar{n}_\mathrm{ph} m c^2 k^2}{16} N\thickspace , \label{eq:SNR_coh} \\ 
    \mathrm{SNR_{incoh}} &= \frac{P_\mathrm{signal} - P_\mathrm{noise}}{\sigma_{P_\mathrm{noise}}} = \frac{\bar{n}_\mathrm{ph} m c^2}{16} \sqrt{N} -1 \approx \frac{\bar{n}_\mathrm{ph} m c^2}{16} \sqrt{N} \thickspace .\label{eq:SNR_incoh}
\end{align}
The SNR in the incoherent protocol does not depend on the signal reduction factor $k$ as no rectification is applied.
For our experimental parameters of $c \approx \SI{30}{\percent}$, $\bar{n}_\mathrm{ph} = 0.057$, $m = 4000$ and $k \approx 0.36$ the expected SNR scaling can be calculated and compared to the experimentally observed scaling in Fig. 3 c of the main text.
The respective values are shown  in Supplementary Tab. \ref{tab:SNR_scaling}.
The theoretical model and the experimental values are in good agreement, confirming the theoretical derivation.

\begin{table}[]
\centering
\caption{Theoretical and experimental SNR scaling.}
\label{tab:SNR_scaling}
\begin{tabular}{cccccccc}
    \hline
    Experiment & Averaging Type &Theoretical scaling & Experimental Scaling \\ \hline
    No Rectification & incoherent &  $\mathrm{SNR_{incoh}} = 1.28 \sqrt{N}$ & $\mathrm{SNR_{incoh}} = \SI{0.971(2)}{} \sqrt{N}$  \\ 
    \textit{in situ} Rectification & coherent & $\mathrm{SNR_{coh}} = 0.16 N$ & $\mathrm{SNR_{coh}} = \SI{0.135(2)}{} N$ \\ 
    \textit{ex situ} Rectification & coherent & $\mathrm{SNR_{coh}} = 0.16 N$ & $\mathrm{SNR_{coh}} = \SI{0.111(1)}{} N$\\  \hline 
\end{tabular}
\end{table}

\section*{Supplementary Note 7: Comparison of different sensing protocols}
This section aims to give an overview about the choice of experimental sequences for the detection of NV-NMR signals as shown in Fig. 4 of the main text by means of a comparison of the SNR in the PSD.
To focus on the influence of the detection protocols, a uniform ensemble of identical $\SI{10}{nm}$ deep NV centers is assumed, where the number of NVs $n_\mathrm{NV}$ in the detection window is set by adjusting the diameter of the confocal laser spot and the number of emitted photons scales linearly with the number of NVs.
A typical contrast $c$ of $\SI{30}{\percent}$ ($\SI{5}{\percent}$) is assumed for the single (ensemble) NVs.
The number of points per trace $m = 4000$ is kept constant for all protocols.
The reference SNR is given by the detection of statistical polarization with a single NV center without rectification (\textit{Qdyne}, rel. $\mathrm{SNR_{ref}} = 1$, gray line).
After 5 minutes of measurement time, the SNR of the single NV can be improved approximately 4 times (see Fig. 3 c) by using the rectification protocols in our experimental setup (dashed gray line in Fig. 4).
In general, this assumes an identical initialization fidelity, sequence duration and rectification fidelity, as in our experimental results shown in Fig. 3 c of the main text. 
Using an ensemble of NV centers, the contrast is reduced by a factor of 6, reducing the SNR by $\frac{1}{36}$ (see Eq. \ref{eq:SNR_coh} and Eq. \ref{eq:SNR_incoh}).
Detection of statistical polarization using an ensemble of NV centers without rectification (blue dashed line) gains no improvement with increasing $n_\mathrm{NV}$ as discussed in the main text, yielding $\mathrm{SNR}/\mathrm{SNR_{ref}} = 1/36$.
By using the rectification protocol, the ensemble SNR (bright blue line) is increased linearly with $n_\mathrm{NV}$ as $\bar{n}_\mathrm{ph} \propto n_\mathrm{NV}$ and follows $\mathrm{SNR}/\mathrm{SNR_{ref}} = 4/36 \thickspace n_\mathrm{NV} $, where we assumed identical rectification efficiency (i.e. the same reduction factor $k$) as in the single NV center case.
In correlation spectroscopy sequences (green dashed line, see e.g. Refs. \cite{liu2022surface,aslam2017}) the sampling of a single trace takes $(m+1)/2$ times longer, thus reducing the number of available traces $N$ by the same factor.
Since the SNR in coherent averaging scales linear with $N$, the SNR is reduced by a factor of $2/(4000+1)$.
Correlation spectroscopy does not suffer from rectification infidelities compared to the single NV rectified case, leading to $\mathrm{SNR}/\mathrm{SNR_{ref}} = \frac{4}{36 \cdot 0.4^2 \cdot 2000}n_\mathrm{NV} $.
The thermal polarization ($\frac{\Delta p}{p} = \tanh{(\hbar \omega/2 k_B T)} \approx \SI{e-5}{}$, $\hbar$: Planck's constant, $\omega$: Zeeman splitting of proton spins, $T$: temperature, $k_B$: Boltzmann constant) is significantly smaller than the statistical polarization ($\frac{\Delta p}{p} = \frac{1}{\sqrt{n_\mathrm{sp}}} = \frac{1}{\sqrt{\rho V_\mathrm{det} }} \approx \SI{3e-3}{}$, $\rho$: proton spin density, $\approx \SI{100}{Mol \per L}$ in oil \cite{aslam2017}, $V_\mathrm{det}$: detection volume, hemisphere with radius of the depth of the NV center).
The detection of thermal polarization using coherent averaging techniques, such as CASR (orange dashed line, see e.g. Refs \cite{grafensteinCoherent2025,glennCASR2018}) suffers from the 300 times lower thermal polarization, which reduces the obtained signal contrast $c$ linearly.
However, avoiding the rectification infidelities, the SNR obtained by CASR protocols is not reduced by the reduction factor $k$, leading to $\mathrm{SNR}/\mathrm{SNR_{ref}} = \frac{4}{36 \cdot 0.4^2 \cdot 300^2}n_\mathrm{NV} $.

\section*{Supplementary Note 8: Variables}
Variables used in the main text and in the supplemental material are given in Supplementary Table \ref{tab:variables}.

\begin{table}[]
\centering
\caption{Table of used variables.}
\label{tab:variables}
\begin{tabular}{cccccccc}
    \hline
    Variable & Description \\ \hline
    $N$ & Number of averages (index $i = 0, ..., N-1$)\\
    $m$ & Number of sampling points in each photon time trace (index $j = 0, ..., m-1$) \\
    $T$ & Total measurement time \\
    $t_\mathrm{seq}$ & Total duration of sequence \\
    $n_\mathrm{SSR}$ & Repetition number of single shot readout \\
    $t_\mathrm{meas, raw}$ & Raw measurement time without experimental overhead through refocussing \\
    $t_\mathrm{meas, total}$ & Total measurement time including experimental overhead\\ \hline
    $B_0$ & Magnetic field \\
    $B_\mathrm{signal}$ & Orthogonal magnetic field amplitude of the applied RF \\
    $f$ & Applied target signal frequency \\
    $f_0$ & Detected target signal frequency \\
    $N_\mathrm{NV}$ & Number of NVs \\
    $d$ & Depth of NV center \\ \hline
    $\mu$ & Nuclear magnetic moment \\
    $n_\mathrm{sp}$ & Number of nuclear spins in the detection volume \\
    $\Delta p/p$ & Relative polarization \\
    $T_1$ & Longitudinal relaxation time of the nitrogen memory qubit \\ \hline
    $M_i$ & Single photon time trace \\
    $s_i$ & Signal amplitude in $M_i$ \\
    $x_i$ & Noise amplitude in $M_i$ \\
    $\phi_i$ & Initial phase of the target signal \\
    $\phi_{i,j}$ & Phase of the target signal at measurement point $j$ \\ \hline
    $F$ & Rectification fidelity \\
    $k$ & Signal reduction factor induced by rectification infidelity \\
    $\bar{n}_\mathrm{ph}$ & Average number of photons per readout \\
    $\alpha$ & Interaction strength between electron spin and target signal \\
    $\theta$ & Acquired phase during single sensing step \\ 
    $P_\mathrm{Signal}$ & Power of the signal in the PSD \\
    $P_\mathrm{Noise}$ & Power of the noise in the PSD \\
    $\sigma_{P_\mathrm{Noise}}$ & Standard deviation of the power of the noise in the PSD \\ \hline
    $\lambda$ & Wavelength of MW\\
    $c$ & Speed of light \\
    $\epsilon_\mathrm{eff}$ & Effective permittivity \\
    $L_\mathrm{res}, W_\mathrm{res}, L_\mathrm{p}, W_\mathrm{p}, L_\mathrm{tap}, L_\mathrm{W}, W_\mathrm{W}$ & Geometry parameters of the MW resonator\\ \hline 
\end{tabular}
\end{table}


\newpage
\bibliographystyle{naturemag}
\bibliography{references.bib}